\documentclass[apj]{emulateapj}

\usepackage{natbib}
\shorttitle{Secondary Eclipse Photometry of WASP-48b and HAT-P-23\lowercase{b}}
\shortauthors{O'Rourke et al.}

\begin{document}

\title{Warm \textit{Spitzer} and Palomar Near-IR Secondary Eclipse Photometry of \\ Two Hot Jupiters: WASP-48\lowercase{b} and HAT-P-23\lowercase{b}}
\author{Joseph G. O'Rourke\altaffilmark{1}, Heather A. Knutson\altaffilmark{1}, Ming Zhao\altaffilmark{2,3}, Jonathan J. Fortney\altaffilmark{4}, Adam Burrows\altaffilmark{5}, Eric Agol\altaffilmark{6}, Drake Deming\altaffilmark{7}, Jean-Michel D\'{e}sert\altaffilmark{1,8}, Andrew W. Howard\altaffilmark{9}, Nikole K. Lewis\altaffilmark{10,13}, Adam P. Showman\altaffilmark{11}, \\ Kamen O. Todorov\altaffilmark{12}}

\altaffiltext{1}{Division of Geological and Planetary Sciences, California Institute of Technology, Pasadena, CA 91125, USA}
\altaffiltext{2}{Department of Astronomy and Astrophysics, 525 Davey Laboratory, The Pennsylvania State University, University Park, PA 16802, USA}
\altaffiltext{3}{Center for Exoplanets and Habitable Worlds, 525 Davey Laboratory, The Pennsylvania State University, University Park, PA 16802, USA}
\altaffiltext{4}{Department of Astronomy and Astrophysics, University of California, Santa Cruz, CA 95064, USA}
\altaffiltext{5}{Department of Astrophysical Sciences, Princeton University, Princeton, NJ 05844, USA}
\altaffiltext{6}{Department of Astronomy, University of Washington, Box 351580, Seattle, WA 98195, USA}
\altaffiltext{7}{Department of Astronomy, University of Maryland, College Park, MD 20742, USA}
\altaffiltext{8}{CASA, Department of Astrophysical and Planetary Sciences, University of Colorado, Boulder, CO 80309, USA}
\altaffiltext{9}{Institute for Astronomy, University of Hawaii, 2680 Woodlawn Drive, Honolulu, HI 96822, USA}
\altaffiltext{10}{Department of Earth, Atmospheric and Planetary Sciences, Massachusetts Institute of Technology, Cambridge, MA 02139, USA}
\altaffiltext{11}{Lunar and Planetary Laboratory, University of Arizona, Tucson, AZ 85721, USA}
\altaffiltext{12}{Institute for Astronomy, ETH Z\"{u}rich, Wolfgang-Pauli-Strasse 27, 8093 Z\"{u}rich, Switzerland}
\altaffiltext{13}{Sagan Fellow}

\begin{abstract}
We report secondary eclipse photometry of two hot Jupiters, WASP-48b and HAT-P-23b, at 3.6 and 4.5~$\mu$m taken with the InfraRed Array Camera aboard the \textit{Spitzer Space Telescope} during the warm \textit{Spitzer} mission and in the $H$ and $K_S$ bands with the Wide Field IR Camera at the Palomar 200-inch Hale Telescope. WASP-48b and HAT-P-23b are Jupiter-mass and twice Jupiter-mass objects orbiting an old, slightly evolved F star and an early G dwarf star, respectively. In the $H$, $K_S$, 3.6~$\mu$m, and 4.5~$\mu$m bands, respectively, we measure secondary eclipse depths of 0.047\% $\pm$ 0.016\%, 0.109\% $\pm$ 0.027\%, 0.176\% $\pm$ 0.013\%, and 0.214\% $\pm$ 0.020\% for WASP-48b. In the $K_S$, 3.6~$\mu$m, and 4.5~$\mu$m bands, respectively, we measure secondary eclipse depths of 0.234\% $\pm$ 0.046\%, 0.248\% $\pm$ 0.019\%, and 0.309\% $\pm$ 0.026\% for HAT-P-23b. For WASP-48b and HAT-P-23b, respectively, we measure delays of 2.6 $\pm$ 3.9 minutes and 4.0 $\pm$ 2.4~minutes relative to the predicted times of secondary eclipse for circular orbits, placing 2$\sigma$ upper limits on $|e\cos\omega|$ of 0.0053 and 0.0080, both of which are consistent with circular orbits. The dayside emission spectra of these planets are well-described by blackbodies with effective temperatures {of 2158 $\pm$ 100~K (WASP-48b) and 2154 $\pm$ 90~K (HAT-P-23b)}, corresponding to moderate recirculation in the zero albedo case. Our measured eclipse depths are also consistent with one-dimensional radiative transfer models featuring varying degrees of recirculation and weak thermal inversions or no inversions at all. We discuss how the absence of strong temperature inversions on these planets may be related to the activity levels and metallicities of their host stars.

\noindent\textit{Key words:} eclipses -- planetary systems -- stars: individual (\objectname{WASP-48}) -- stars: individual (\objectname{HAT-P-23}) -- techniques: photometric
\end{abstract}

\section{Introduction}

Hot Jupiters are the most easily detectible class of exoplanets using either the radial velocity or transit techniques. These massive gas giants, unlike anything in our Solar System, are locked in scorching orbits around their parent stars, with orbital periods of a few days and very high equilibrium temperatures---in some cases over 3000~K. By studying the properties of their unusual atmospheres, we can gain important clues to their formation histories and the effects that their exotic environments might have on their atmospheric chemistry.

We can determine the properties of the dayside atmospheres of hot Jupiters using the secondary eclipse technique. During a secondary eclipse, the thermal emission from the planet's dayside is blocked as the planet moves behind its parent star. With observations at several wavelengths, we can construct a rough dayside emission spectrum for the exoplanet \citep[e.g.,][]{Charbonneau2008, Knutson2008}. Since the first measurements of thermal emission from hot Jupiters \citep{Deming2005,Charbonneau2005}, the \textit{Spitzer Space Telescope} has observed the secondary eclipses of nearly fifty extrasolar planets. Today, observations continue during \textit{Spitzer}'s extended warm mission; although the last of the telescope's cryogen was exhausted in 2009, the Infrared Array Camera (IRAC; \citealp{Fazio2004}) continues operating in its 3.6 and 4.5~$\mu$m bands.

Ground-based observations of secondary eclipses in the near-IR provide an important complement to these \textit{Spitzer} observations \citep[e.g.,][]{Rogers2009,Sing2009,Gibson2010,Croll2010,Croll2011,Caceres2011,DeMooij2011,Zhao2012,Zhao2012b}. While the bright sky background at $\sim$3-4~$\mu$m typically prevents observations of secondary eclipses in the $L$ and $M$ bands, many hot Jupiters are easily accessible at shorter wavelengths. The planet/star flux ratio, which determines the secondary eclipse depth, decreases at shorter wavelengths, so the hottest planets ($>$2000~K) with their correspondingly high near-IR fluxes generally make the most favorable targets for ground-based eclipse observations. The primary challenge in obtaining reliable ground-based observations of secondary eclipses lies in removing the time-varying telluric and instrumental effects, which often limit the achieved precision to a factor of a few above the photon noise limit \citep[e.g.,][]{Croll2010,Croll2011,Zhao2012,Zhao2012b}. Ultimately, observations in \textit{both} the near- and mid-IR from the ground and space place the tightest constraints on the characteristics of the atmospheres of hot Jupiters.

Observations of secondary eclipses provide invaluable information about the properties of these systems. First, knowing the precise timing of the secondary eclipse aids in constraining the eccentricity of a planet's orbit \citep{Charbonneau2005}, which provides important clues as to the efficiency of orbital circularization for close-in planets. Furthermore, orbital eccentricity may determine whether tidal heating from ongoing orbital circularization can explain the inflated radii of certain planets \citep[e.g.,][]{Bodenheimer2001}.

Ongoing surveys of the emission spectra of hot Jupiters indicate that some host temperature inversions in their upper atmospheres while others do not {\citep[e.g.,][]{Knutson2008,Burrows2008,Fortney2008,Barman2008,Madhusudhan2009,Fressin2010}.} These inversions are likely caused by the presence of a high-altitude absorber, but the nature of this absorber is currently debated. TiO has been proposed as one possible candidate \citep{Hubeny2003,Fortney2008}, but at least one hot Jupiter appears to have an inversion even though its dayside is too cold for TiO to exist in the gas phase \citep{Machalek2008}. Moreover, even if the dayside is hotter than the condensation temperature for TiO, it may still be lost to settling or cold traps on the nightside and in the deep interior \citep{Showman2009,Spiegel2009}. {For example, WASP-19b has a dayside that is hot enough for gas-phase TiO, but appears to lack both an inversion \citep{Anderson2013} and TiO features in its transmission spectrum \citep{Huitson2013}.}

Because the stellar flux incident from above is much greater than the internal luminosity from below, the atmospheres of hot Jupiters tend to be stably stratified. Scaling arguments and general circulation model (GCM) simulations show that, despite the stable stratification, vigorous vertical mixing occurs at and above photospheric levels \citep[e.g.,][]{Showman2002,Showman2008,Showman2009,DobbsDixon2008,Heng2011,Rauscher2012,Parmentier2013a}. Nevertheless, at deeper levels of $\sim$100~bars, the vertical mixing rate may be too low to prevent the loss of TiO particles by downward settling. Alternative absorbers include sulfur-containing compounds \citep{Zahnle2009}. There is also empirical evidence for a correlation between hot Jupiter emission spectra and stellar activity \citep{Knutson2010}, indicating that the increased UV radiation from more active stars may destroy the atmospheric absorber(s) that create thermal inversions. Variations in elemental abundances, such as elevated C/O ratios, may also explain the lack of strong inversions in some planets \citep{Madhusudhan2011,Madhusudhan2012}.

The secondary eclipse of a hot Jupiter can also be used to constrain the efficiency of energy transport from the dayside to the nightside. \citet{Cowan2011a} examine the available secondary eclipse data for hot Jupiters and conclude that planets with equilibrium temperatures above $\sim$2400~K transport very little of the incident energy to their nightsides, whereas hot Jupiters with lower equilibrium temperatures show a broad range in transportation efficiencies and/or dayside albedos. By building up a large sample of hot Jupiters with well-characterized dayside emission spectra, we can search for correlations between the efficiency of energy transport or the presence of temperature inversions and other properties of the system (e.g., stellar metallicity, activity, or planetary mass) that might allow us to discriminate between various physical models for these processes.  

In this paper, we present multi-band secondary eclipse observations for the transiting hot Jupiters WASP-48b and HAT-P-23b. We observe these secondary eclipses in the $H$ and $K_S$ bands with the Wide Field IR Camera at the Palomar 200-inch Hale Telescope \citep{Wilson2003} and in the 3.6 and 4.5~$\mu$m \textit{Spitzer} bands. WASP-48b is a 0.98$M_J$ planet orbiting with a period of 2.14~days at 0.034~AU around a slightly evolved 1.19$M_\sun$ star \citep{Enoch2011}. WASP-48 has a large radius (1.75$R_\odot$) and a very short rotational period ($\sim$7~days) for its $\sim$8~Gyr age, possibly indicating that it was spun up through tidal interactions with WASP-48b \citep{Enoch2011}. HAT-P-23b is a 2.09$M_J$ planet with an inflated radius of $\sim$1.37$R_J$, in a prograde, aligned orbit with a period of 1.21~days at 0.023~AU around an early G dwarf star \citep{Bakos2011,Moutou2011}. Both planets have zero-albedo equilibrium temperatures above 2000~K, which makes them favorable targets for ground-based observations. In Section~\ref{sec:obs}, we describe our observations and our procedure for extracting time-series photometry from the collected images. In Section~\ref{sec:analysis}, we calculate the timing and depth of the secondary eclipses from the retrieved photometry. In Section~\ref{sec:diss}, we discuss how the timing of the eclipses constrains the orbital eccentricities of the planets, and we compare our eclipse depths to various models for the thermal emission spectra of the planetary daysides.

\section{Observations and Photometry}
\label{sec:obs}

\subsection{Secondary Eclipse Observations with Spitzer}

We observed both planets in secondary eclipse using the InfraRed Array Camera (IRAC) on board the \textit{Spitzer Space Telescope}. For all observations, we used the IRAC full array mode, which produced images of 256 $\times$ 256 pixels (5$\farcm$2 $\times$ 5$\farcm$2), with 12.0~s integration times over 7.9~hours. For WASP-48b, we obtained 2167 images for both the 3.6 and 4.5 $\mu$m bands on UT 2011 August 09 and UT 2011 September 07, respectively. For HAT-P-23b, we obtained 2166 images in the same bands on UT 2011 November 30 and UT 2011 December 01. WASP-48 and HAT-P-23 have $K_S$ band magnitudes of 10.4 and 10.8, respectively.

We extract photometry from the basic calibrated data (BCD) files produced by versions S.18.18.0 (for WASP-48b, 3.6~$\mu$m) and S.19.0.0 (for HAT-P-23b, 3.6 and 4.5~$\mu$m, and WASP-48b, 4.5~$\mu$m) of the IRAC pipeline. The pipeline dark-subtracted, flat-fielded, linearized, and flux-calibrated the images. For each image, we extract the UTC-based Barycentric Julian Date (BJD$_{UTC}$) from the FITS header (keyword BMJD$\textunderscore$OBS). We transform these time stamps into the Barycentric Julian Date based on the Terrestrial Time standard (BJD$_{TT}$) using the conversion at the time of our observations, BJD$_{TT}$ $\approx$ BJD$_{UTC}$ + 66.184~s \citep{Eastman2010}. We prefer the continuous BJD$_{TT}$ standard because leap seconds are occasionally added to the BJD$_{UTC}$ standard.

We correct for transient ``hot pixels" within a 20$\times$20 pixel box centered on our target star. In each frame, we compare the intensity of each individual pixel within the box to its median value in the 10 preceding and 10 following frames. If the intensity of a pixel varies by $>$3$\sigma$ from this local median value, we replace it with that value. We corrected 0.18\% and 0.25\% of pixels in the 3.6 and 4.5~$\mu$m bands, respectively, for WASP-48b, and 0.16\% and 0.20\% of pixels for HAT-P-23b.

To find the center of the stellar point spread function (PSF), we use the flux-weighted centroid method. That is, we first calculate the flux-weighted centroid within 4.0 pixels of the approximate position of the target star. Then, we re-calculate the centroid, using the first centroid as our new approximate position. We iterate this procedure a total of 10 times and use the final centroid as the stellar position. Alternatively, we could fit a two-dimensional Gaussian function to the stellar image, but prior experience with extracting time series photometry from \textit{Spitzer} data shows that the flux-weighted centroid method is either equivalent or superior \citep[e.g.,][]{Beerer2011,Knutson2012,Lewis2013}. The \textit{x} and \textit{y} coordinates of our target stars changed by 0.26 pixels or less during all of our \textit{Spitzer} observations.

We extract photometry using two methods. First, we perform aperture photometry with the IDL routine \textit{aper}, using circular apertures with radii ranging from 2.5 to 5.0 pixels in 0.1 pixel increments. We choose aperture radii that yield the lowest scatter in the residuals of the eventual light curves, but we also confirm that eclipse depths and timings remain consistent for the entire range of aperture settings. Second, we use a time-varying aperture radius based on the noise-pixel parameter, which is proportional to the square of the full-width-half-maximum of the stellar PSF \citep{Mighell2005,Knutson2012,Lewis2013}. According to Section 2.2.2 of the \textit{Spitzer/IRAC} instrument handbook, the noise-pixel parameter is defined as: 
\begin{equation}
\bar\beta = \frac{(\sum_iI_i)^2}{\sum_iI_i^2},
\end{equation}
where $I_i$ is the intensity recorded by the $i$th pixel. We include all pixels that are located at least partially within a 4.0 pixel radius of the stellar centroid, which we found minimizes the scatter in the residuals from our best-fit solution. We extract photometry from each image using a circular aperture of radius
\begin{equation}
r = a_0\sqrt{\bar\beta} + a_1,
\end{equation}
where $a_0$ and $a_1$ are constants that we vary in steps of 0.1. We found that photometry based on the noise pixel {decreased the scatter in the residuals of our ultimate light curve by $\sim$5\%} for the secondary eclipse of WASP-48b in the 3.6~$\mu$m band, with a median radius of $\sim$2.5 pixels for the photometric aperture, {although we obtained consistent eclipse depths and timings using fixed radii}. This time-varying approach did not improve the fits for the other secondary eclipses, for which we used circular apertures with fixed radii of 2.8 pixels (WASP-48b, 4.5~$\mu$m) and 3.0 pixels (HAT-P-23b, 3.6~$\mu$m and 4.5~$\mu$m). We estimate the background using the 3-$\sigma$ clipped mean within circular sky annuli with inner and outer radii, respectively, of 20.0 and 30.0 pixels (WASP-48b) and 30.0 and 50.0 pixels (HAT-P-23b). There were no visible bright stars within these annuli. Using different annuli yields consistent results, but with higher scatter in the residuals of the eventual light curves. To estimate the Poisson noise limit for the photometry, we converted the pixel intensities to electron counts from MJy/sr using the conversion factors in the FITS headers.

\subsection{Secondary Eclipse Observations at Palomar}

\begin{figure}
\noindent\includegraphics[width=20pc]{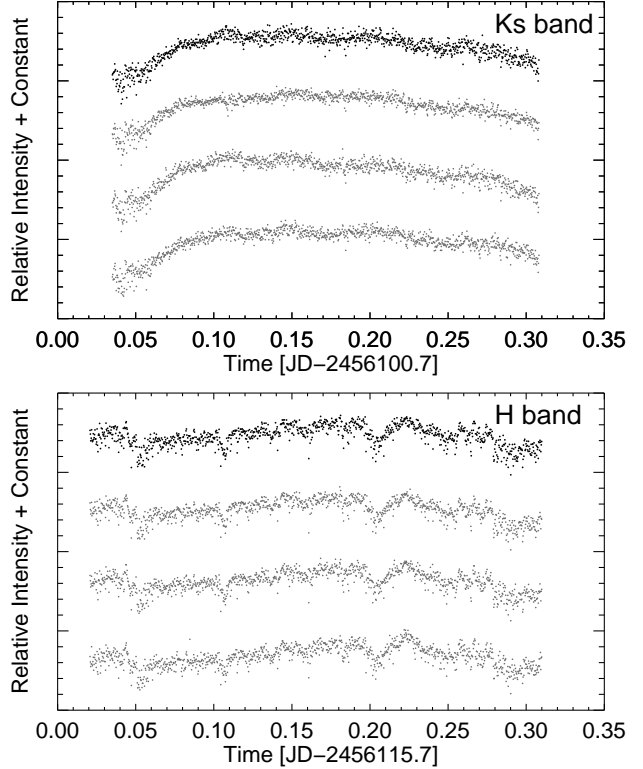}
\caption{Raw photometry vs. time WASP-48b (black, upper curve) and calibration stars (grey, lower curves) in the $K_S$ and $H$ bands.} \label{fig:wirc_raw48}
\end{figure}

\begin{figure}
\noindent\includegraphics[width=20pc]{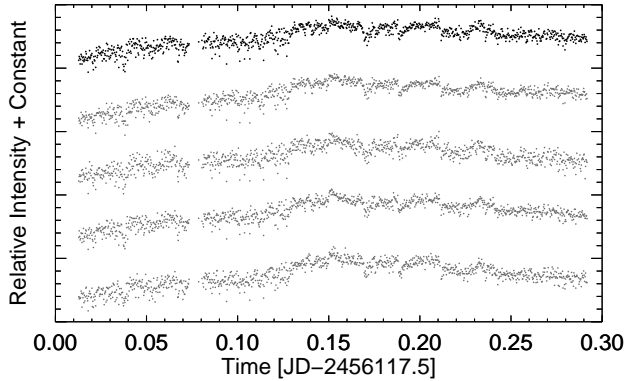}
\caption{Raw photometry vs. time for HAT-P-23b (black, upper curve) and four calibration stars (grey, lower curves) in the $K_S$ band.} \label{fig:wirc_raw23}
\end{figure}

We observed secondary eclipses of HAT-P-23b in the $K_S$ band and WASP-48b in the $H$ and $K_S$ bands using the Wide field IR Camera (WIRC) instrument on the Palomar 200-inch Hale telescope. The WIRC instrument has a 2048 $\times$ 2048 detector with a field of view of 8\farcm7 $\times$ 8\farcm7 and a pixel scale of 0\farcs2487. The seeing was variable during each night but generally $\lesssim1\farcs5$. 

To avoid saturating the detector and to minimize systematic errors resulting from variations in intra-pixel sensitivity, we defocused the telescope to a FWHM of $\sim$2$\farcs5$-$3\arcsec$. We used the new active guiding scheme for WIRC \citep{Zhao2012} to improve the tracking of the telescope and thus mitigate the effects of imperfect flat-fielding and inter-pixel variations in the detector. To minimize these systematics, we also refrained from dithering during all observations. We positioned the telescope so that the target and calibration stars were located as far as possible from bad pixels. The $x$ and $y$ coordinates of the target star varied smoothly throughout the night by less than 3.7 and 2.5 pixels in the $K_S$ band and 5.4 and 5.3 pixels in the $H$ band (WASP-48b), respectively, and 5.9 and 4.5 pixels in the $K_S$ band (HAT-P-23b), which is $\sim$3-5 times better than the pointing stability before the work of \citet{Zhao2012}.

Our images of WASP-48b in the $K_S$ band were obtained on UT 2012 June 22 with a total duration of 6.6 hours, yielding 1094 images with an exposure time of 8~s. Our observations of WASP-48b in the $H$ band were obtained on UT 2012 July 07 with a total duration of 6.9 hours, yielding 1064 images with an exposure time of 10~s. Our images of HAT-P-23b in the $K_S$ band were obtained on UT 2012 July 09 with a total duration of 6.7 hours, yielding 1037 images with an exposure time of 9~s. The airmass ranged from 1.08-1.35, 1.08-1.47, and 1.04-1.53 during the nights, respectively. 

We extracted time-series photometry for the target stars and calibration stars in the collected images. First, we subtracted averaged dark frames from all images. We combined twilight flats into master flat fields by normalizing each individual flat to unity and then taking the median for each night. For the $K_S$ and $H$ band observations of WASP-48b, respectively, we selected three calibration stars with median fluxes ranging from $\sim$0.8-1.1 and $\sim$0.2-0.9 times that of WASP-48. For HAT-P-23b, we selected four calibration stars with median fluxes ranging from $\sim$0.4-0.8 times that of HAT-P-23. Brighter stars in the fields were excluded because their fluxes exceeded the linearity regime or saturated the detector. Faint stars in the fields had insufficient signal-to-noise and were thus ignored. Still other stars were neglected if their inclusion into the analysis resulted in an increased scatter in the best-fit residuals. We corrected for transient ``hot pixels" around each star as for the warm \textit{Spitzer} data.

To determine the position of each star, we calculated the flux-weighted centroid within 20.0 pixels of the approximate center of the star. We allowed each stellar position to vary independently. For these data, we cannot obtain accurate position estimates using a two-dimensional Gaussian fit because defocusing the telescope makes the stellar PSFs highly irregular, with elliptical shapes and uneven distributions of flux. We corrected the time stamp of each image to the mid-exposure time and then converted the time into the BJD$_{TT}$ standard using the routines of \citet{Eastman2010}. We extract conversion factors from the FITS headers to estimate the Poisson noise limits for the photometry.

We performed aperture photometry on each star using circular apertures with fixed radii ranging from 8.0-20.0 pixels in increments of 0.5 pixels. Radii of 11.5 pixels (WASP-48b, $K_S$ band), 15.0 pixels (WASP-48b, $H$ band), and 11.0 pixels (HAT-P-23b, $K_S$ band) ultimately produced the best fits. We used the same photometric apertures for the target and calibration stars in each set of observations; varying the aperture size from star to star during the analysis degraded the quality of the resulting light curves because the shapes of the stellar PSFs were not constant during the night. We used sky annuli with inner and outer radii of 30.0 and 55.0 pixels, respectively. We again used the 3-$\sigma$ clipped mean as our background value and checked to make sure that there were not any bright stars within the annuli. We selected this annulus size and range to minimize the RMS scatter in the residuals residuals from the final fits, but different annuli yield consistent results for the eclipse depths and times. Figures~\ref{fig:wirc_raw48} and~\ref{fig:wirc_raw23} show the raw photometry that we extracted for WASP-48b, HAT-P-23b, and calibration stars during all three nights.

\section{Data Analysis}
\label{sec:analysis}

\subsection{Ephemerides}

The data analysis routines require precise ephemerides for the systems. In particular, we need to know the orbital period, the ratio of the stellar and planetary radii, the orbital inclination, the orbital semi-major axis, and the timing of at least one eclipse or transit. Accounting for the orbit-crossing time of light and assuming circular orbits, secondary eclipses would occur at a phase of 0.5002 for both planets. We use the ephemerides from the discovery papers for  HAT-P-23b \citep{Bakos2011} and WASP-48b \citep{Enoch2011}. \citet{Sada2012} observed a more recent transit of WASP-48b and calculated different values for its inclination and semi-major axis, which are tightly correlated. We obtain consistent results with both sets of ephemerides. However, we use the original ephemerides from \citet{Enoch2011} because they ultimately yield lower formal errors. 

\subsection{Analysis of Spitzer Data}

\begin{figure}
\noindent\includegraphics[width=20pc]{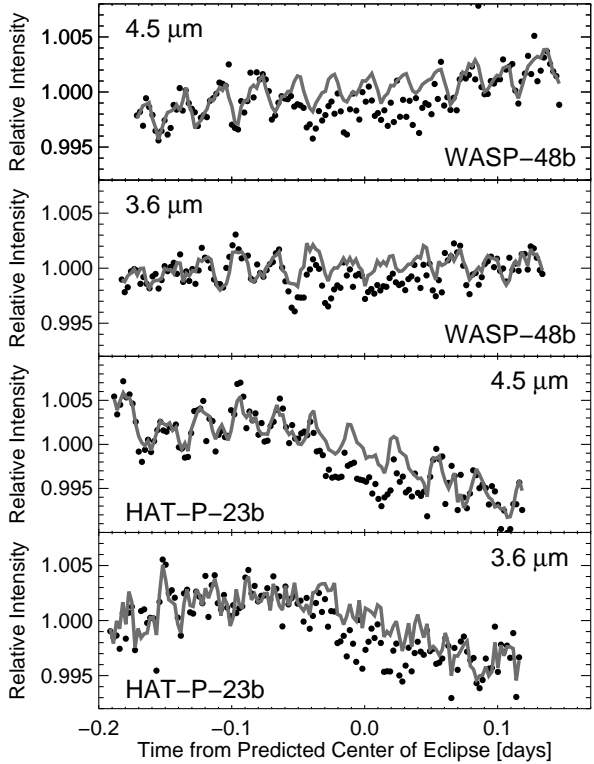}
\caption{Photometry in the Spitzer 3.6~$\mu$m and 4.5~$\mu$m bands vs. time from the predicted centers of secondary eclipse for WASP-48b and HAT-P-23b. The functions used to correct for intrapixel sensitivity and linear trends in time are overplotted in grey. Data are binned in 3 minute intervals.} \label{fig:Spitzer_raw}
\end{figure}

We use standard methods to analyze the data from warm \textit{Spitzer} \citep[e.g.,][]{Beerer2011,Knutson2012,Lewis2013,Baskin2013,Todorov2013}. We initially discard points in our light curve that suffered uncorrected particle hits within our science aperture or significant spatial displacement. Specifically, we discard points in which the measured flux or the \textit{x} or \textit{y} position vary by $>$$3\sigma$ from the median values in the adjacent 20 frames in the time series. In total, we excised 23 (1.1\%) and 39 (1.8\%) frames from the 3.6 and 4.5~$\mu$m observations for WASP-48b, respectively, and 23 (1.1\%) and 30 (1.4\%) frames for HAT-P-23b. 

The background fluxes for these data sets exhibit a ramp-like behavior and intermittent variation between three distinct levels, particularly in the 3.6~$\mu$m band \citep[e.g.,][]{Beerer2011,Deming2011}. Also, the position of the target star on the detector does not stabilize for $\sim$15-30 minutes. To counter these effects, we trim the number of frames from the beginning of each time series that minimizes the RMS of the residuals in our best-fit solution: 126 (HAT-P-23b, 3.6~$\mu$m), 131 (HAT-P-23b, 4.5~$\mu$m), 69 (WASP-48b, 3.6~$\mu$m), and 60 (WASP-48b, 4.5~$\mu$m).

The total flux measured within our photometric aperture varies due to a well-known intrapixel sensitivity effect \citep[e.g.,][]{Charbonneau2005,Charbonneau2008}. We first normalize our light curve so that the median of the out-of-eclipse flux values is equal to unity. Then, to correct for the systematic effects, we fit the data to ``decorrelation functions" involving the measured \textit{x} and \textit{y} positions of the star, along with a term that is linear in time to reduce correlated noise. Our most general decorrelation function is 
\begin{equation}
F(\{c_i\},\bar x,\bar y,t)=c_0 + c_1\bar x + c_2 \bar y + c_3 \bar x^2 + c_4 \bar y^2 + c_5 t, \label{eq:dec}
\end{equation}
where $t$ is the predicted time from the center of the secondary eclipse, $\bar x$ and $\bar y$ are the \textit{x} and \textit{y} positions minus their median values over the time series, and the $\{c_i\}$ are free parameters. Including all of the terms in Eq.~\ref{eq:dec} for the 3.6~$\mu$m band data improved the fits for both targets, relative to fits in which the quadratic terms were excluded, where improvement is defined as a decrease in $\chi^2$ (by 9 for WASP-48b and 2 for HAT-P-23b). We assumed that the measurement error for each of our points was equal to the RMS scatter in the residuals between the photometry and the best-fit eclipse model for each fit, although this is likely an underestimate, as discussed below. Including the $\bar x^2$ and $\bar y^2$ terms in the fits for the 4.5~$\mu$m band data decreased $\chi^2$ by $\le$1 for both targets, so we set $c_3=c_4=0$ in our analysis of these data. We tried including a $\bar x\bar y$ term \citep{Desert2009} and obtained consistent results, but the $\chi^2$ for our fits remained constant or increased. Figure~\ref{fig:Spitzer_raw} shows the raw photometry along with the best-fit decorrelation functions.

The Bayesian information criterion (BIC) is a popular method of model selection \citep{Liddle2007}. We may write
\begin{equation}
\textrm{BIC} = \chi^2 + k \ln(n), \label{eq:bic}
\end{equation}
where $k$ is the number of free parameters in the model and $n$ is the number of observations. Models with a lower BIC are usually preferred. That is, the improvement in the fit achieved by adding an additional parameter to the model must be larger than the penalty imposed by the rightmost term in Eq.~\ref{eq:bic}. For our observations, $\ln (n) \sim 8$. Although we desire a quantitative criterion for which terms to exclude in Eq.~\ref{eq:dec}, the BIC is ill-suited to this purpose \citep{Todorov2013}. In particular, our calculation of $\chi^2$ depends on the intrinsic error in the light curve, $\sigma$, which is certainly larger than the Poisson noise limit because of systematic effects but otherwise undetermined. In any case, we obtain consistent results for the eclipse depths and timing offsets whether or not we include the quadratic terms in Eq.~\ref{eq:dec}.

\begin{figure}
\noindent\includegraphics[width=20pc]{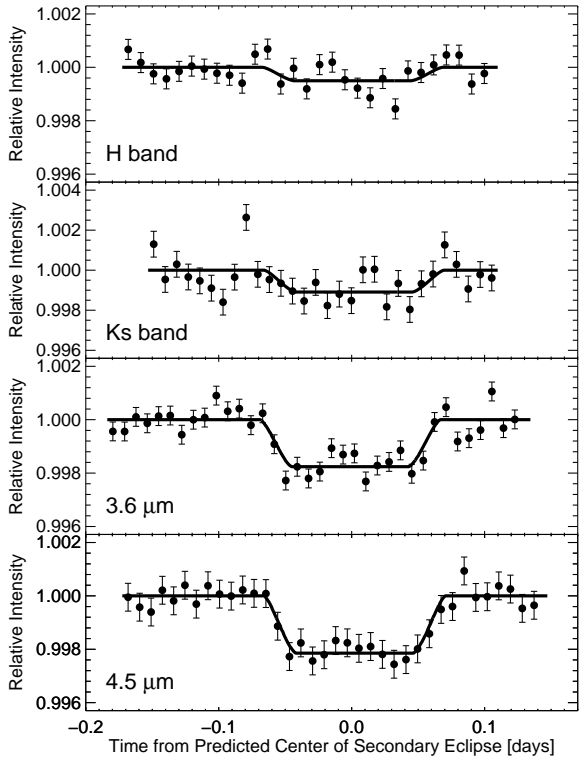}
\caption{Photometry in four wavebands after decorrelation vs. time from the predicted center of secondary eclipse for WASP-48b. Best-fit eclipse light curves are overplotted as solid lines. Data are binned in 12 minute intervals. The lengths of the error bars are based on the RMS scatter in the residuals between the photometry and the best-fit eclipse model over the whole light curve.} \label{fig:fits48}
\end{figure}

\begin{figure}
\noindent\includegraphics[width=20pc]{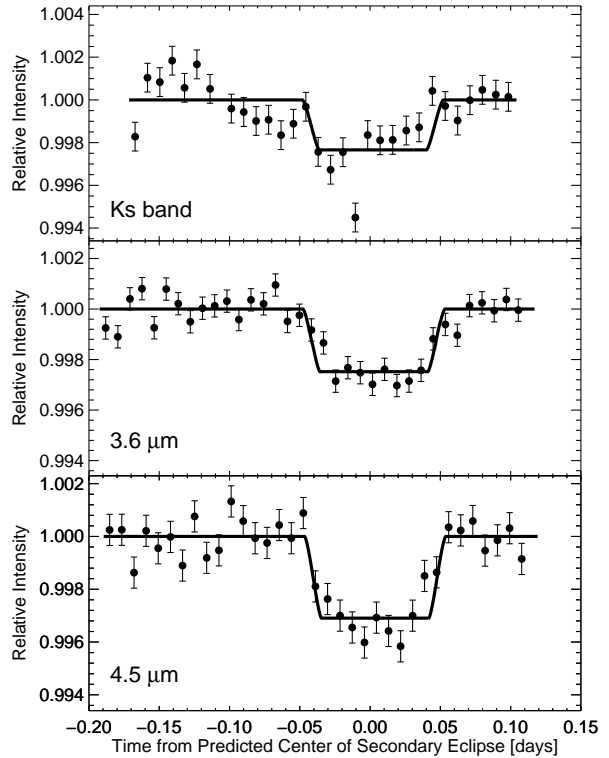}
\caption{Photometry in three wavebands after decorrelation vs. time from the predicted center of secondary eclipse for HAT-P-23b. Best-fit eclipse light curves are overplotted as solid lines. Data are binned in 12 minute intervals. The lengths of the error bars are based on the RMS scatter in the residuals between the photometry and the best-fit eclipse model over the whole light curve.} \label{fig:fits23}
\end{figure}

We simultaneously fit a model for the eclipse, $G(d,\Delta t)$, with two free parameters: the eclipse depth $d$ and a timing offset from the predicted center of eclipse $\Delta t$ \citep{Mandel2002}. Initially, we fit these functions using the Levenberg-Marquardt algorithm \citep{Markwardt2009}. We fit all free parameters simultaneously, but we also tried fitting the decorrelation function and then the eclipse light curve in series; we report the simultaneous fits as our final results, but both methods produce consistent results. Figures~\ref{fig:fits48} and~\ref{fig:fits23} show the decorrelated photometry and the best-fit eclipse models for WASP-48b and HAT-P-23b. The RMS scatters in the residuals are, respectively, 0.390\% and 0.264\% in the 4.5 and 3.6~$\mu$m bands for WASP-48b and 0.442\% and 0.331\% in the same bands for HAT-P-23b. These residual errors are 24.0\% and 14.4\% (WASP-48b) and 14.2\% and 15.0\% (HAT-P-23b) above the Poisson noise limits.

We used two methods to estimate the uncertainties in our calculated eclipse depth and time offsets. First, we used the Markov chain Monte Carlo (MCMC) method \citep{Ford2005,Winn2009} with 10$^7$ steps to fit the parameters, using our first fit as our initial state. We ran independent Markov chains with different initial states and obtained consistent results. The distributions for all parameters were roughly Gaussian, and the eclipse depths and time offsets were neither correlated with each other nor with any parameter in the decorrelation functions. Second, we estimated the contribution to uncertainty from time-correlated noise with the ``prayer-bead" (PB) method \citep{Gillon2009,Carter2009}. That is, we removed the residuals from our best-fit solution, then shifted them in increments of one time step and added them back into the best-fit solution. At each time step, we re-fit the eclipse light curve using the Levenberg-Marquardt algorithm. 

We compared the prayer-bead distribution of eclipse depths and time offsets to the 68\% symmetric confidence intervals from the Markov chains and reported the larger of the two as our formal errors. Both methods returned roughly consistent results, which are summarized in Table~\ref{table}. Specifically, for WASP-48b, the MCMC error for the eclipse depth (0.020\%) was larger than the PB error (0.009\%) in the 4.5~$\mu$m band, whereas the PB and MCMC errors were identical (0.013\%) for the 3.6~$\mu$m band. For HAT-P-23b, the MCMC errors were larger in both the 3.6~$\mu$m band (0.019\% vs. 0.013\%) and the 4.5~$\mu$m band (0.026\% vs. 0.020\%). The PB errors in the eclipse timing were larger for every observation except the 3.6~$\mu$m band observation of HAT-P-23b, for which the MCMC error was 0.1~min larger. 

We measured the timing of the secondary eclipses to within 4.3 and 7.8~min (WASP-48b) and 2.2 and 2.6~min (HAT-P-23b) in the 4.5 and 3.6~$\mu$m bands, respectively. We calculated the predicted timing of each center of eclipse using the orbital period and transit epochs from the discovery papers \citep{Enoch2011,Bakos2011}. We added the uncertainties in the ephemerides in quadrature with our measurement error to obtain our formal errors on the eclipse offset, which are also summarized in Table~\ref{table}. Overall, we have measured mean timing offsets of 2.6 $\pm$ 3.9~min for WASP-48b and 4.0 $\pm$ 2.4~min for HAT-P-23b.

\subsection{Analysis of Palomar Data}

We discarded the final 37 and 21 images from the WASP-48b $K_S$ and $H$ band data, respectively, because the sky background began to increase dramatically near sunrise. We also discarded images if any pixel counts in the photometric apertures exceeded the linearity regime of the detector, or if the total flux for any star varied by $>$3$\sigma$ from the median values in the adjacent 20 frames in the time series. In total, we excised 39, 26, and 13 images for the WASP-48b $K_S$ and $H$ band and the HAT-P-23b $K_S$ band data, respectively, under these two criteria.

Most of the variations in the raw light curves from the ground-based data result from correlated systematics, including changes in seeing and airmass, atmospheric transition, and thermal background \citep[e.g.,][]{Zhao2012}. Fortunately, these systematics have similar effects on the target and calibration stars. For each set of observations, we take the mean of the normalized light curves of the calibration stars to form a single reference light curve. Combining the calibration light curves using the median or flux-weighted mean produced consistent, but inferior, results in the ultimate fits. We also attempted to combine the calibration stars using a weighted average, where the relative weights for each star varied as free parameters in the fit, but this did not improve our ultimate best-fit solution.

\begin{deluxetable*}{cccccc}
\tablecaption{Summary of Secondary Eclipse Results}
\tablehead{\colhead{Planet} & \colhead{Waveband} & \colhead{Depth (\%)} & \colhead{Center of Eclipse (BJD$_{TT}$\tablenotemark{a})} & \colhead{Eclipse Offset\tablenotemark{b} (min)} & \colhead{T$_b$ (K)}}
\startdata
WASP-48b & $H$ & 0.047 $\pm$ 0.016 & 2456115.89713 $\pm$ 0.00292\tablenotemark{c} & 2.6 $\pm$ 3.9\tablenotemark{c} & 2158 $\pm$ 196 \\
WASP-48b & $K_S$ & 0.109 $\pm$ 0.027 & 2456100.89167 $\pm$ 0.00292\tablenotemark{c} & 2.6 $\pm$ 3.9\tablenotemark{c} & 2330 $\pm$ 200 \\
WASP-48b & 3.6 $\mu$m & 0.176 $\pm$ 0.013 & 2455783.63187 $\pm$ 0.00543 & -0.3 $\pm$ 7.9\tablenotemark{d} & 2155 $\pm$ 73 \\
WASP-48b & 4.5 $\mu$m & 0.214 $\pm$ 0.020 & 2455811.50176 $\pm$ 0.00295 & 3.5 $\pm$ 4.5\tablenotemark{d} & 2122 $\pm$ 103 \\ 
HAT-P-23b & $K_S$ & 0.234 $\pm$ 0.046 & 2456117.91288 $\pm$ 0.00271\tablenotemark{c} & 4.0 $\pm$ 2.4\tablenotemark{c} & 2668 $\pm$ 198 \\
HAT-P-23b & 3.6 $\mu$m & 0.248 $\pm$ 0.019 & 2455895.95423 $\pm$ 0.00181 & 2.7 $\pm$ 3.6\tablenotemark{d} & 2128 $\pm$ 74 \\
HAT-P-23b & 4.5 $\mu$m & 0.309 $\pm$ 0.026 & 2455897.16873 $\pm$ 0.00153 & 5.1 $\pm$ 3.3\tablenotemark{d} & 2114 $\pm$ 92  \label{table}
\enddata
\tablenotetext{a}{BJD$_{UTC}$ $\approx$ BJD$_{TT}$ - 66.184~s for our observations before 2012 June 30 and BJD$_{UTC}$ $\approx$ BJD$_{TT}$ - 67.184~s for our observations after 2012 June 30 (WASP-48b in the $H$ band and HAT-P-23b in the $K_S$ band).}
\tablenotetext{b}{We report the delays from the predicted centers of secondary eclipse assuming circular orbits, which would occur at phases of 0.5002 because of the light travel time for both planets.}
\tablenotetext{c}{We used the observed timing of the eclipses in the \textit{Spitzer} bands to fix the eclipse offsets and thus the centers of eclipse for the $H$ and $K_S$ band data. The quoted uncertainties on the centers of these eclipses were calculated from the uncertainties that we reported for the \textit{Spitzer} data, along with the uncertainties in the ephemerides.}
\tablenotetext{d}{The quoted uncertainties in the eclipse offsets for the \textit{Spitzer} data include the measurement error as well as the propagated uncertainties in the orbital periods and the transit epochs from the discovery papers \citep{Bakos2011,Enoch2011}.}
\end{deluxetable*}

We use our averaged calibrator star light curve as part of a decorrelation function to correct for these systematics. We fit the following decorrelation function to the light curve for HAT-P-23b in the $K_S$ band:
\begin{equation}
F(\{c_i\},t)=c_0 + c_1t + c_2L_R, \label{eq:decorr_wirc}
\end{equation}
where $L_R$ is the reference light curve. For the observations of WASP-48b in both the $K_S$ and $H$ wavebands, we obtained consistent results when we simply divided the target light curve by the reference light curve and then fit a decorrelation function that was only linear in time, i.e. with $c_2=0$ in Eq.~\ref{eq:decorr_wirc}. Doing the same for the $K_S$ band observation of HAT-P-23b, however, implausibly caused the eclipse depth to increase by $>$3$\sigma$, along with a $\sim$10\% increase in the residual scatter, so we elected to use the full decorrelation function in Eq.~\ref{eq:decorr_wirc}.

As with the \textit{Spitzer} data, we simultaneously fit a model for the secondary eclipse from \citet{Mandel2002}. However, we elect to fix the time offset to the best-fit value from the analysis of the \textit{Spitzer} data to reduce the number of free parameters in our fit, since our ground-based data are of comparatively lower quality. That is, $\Delta t$ = 0.4 and 9.5~min for WASP-48b and HAT-P-23b, respectively. Changing the eclipse offset by $\pm$1$\sigma$ yields consistent eclipse depths to within $\pm$1$\sigma$. In fact, the best-fit eclipse depth changes by $\le$0.002\% for both observations of WASP-48b. Moreover, we obtain consistent eclipse depths when we constrain the eclipses to have no timing offsets, as predicted for circular orbits. We again first calculate the best-fit values for each parameter using the Levenberg-Marquardt algorithm. We then run Markov chains with 10$^7$ and $5\times10^6$ steps, respectively, for HAT-P-23b and WASP-48b. 

We report the best-fit solutions with minimized $\chi^2$ from the Levenberg-Marquardt algorithm in Table~\ref{table}, but the medians of the distributions from the Markov chains are consistent within $\pm$1$\sigma$. No parameters are strongly correlated. We run our prayer-bead routine on the best-fit solution to provide another measure of uncertainty. The prayer-bead errors for HAT-P-23b are larger than the MCMC errors (0.046\% vs. 0.036\%), which is expected given the large systematic effects seen in Figure~\ref{fig:fits23}, but the MCMC errors are larger for the WASP-48b eclipse depths (0.016\% vs. 0.006\% in the $H$ band and 0.027\% vs. 0.011\% in the $K_S$ band). 

Our results are summarized in Table~\ref{table}. Figures~\ref{fig:fits48} and~\ref{fig:fits23} show our best-fit light curves. The RMS scatters in the best-fit residuals are 0.380\% (WASP-48b, $K_S$ band), 0.222\% (WASP-48b, $H$ band), and 0.393\% (HAT-P-23b, $K_S$ band). These are factors of 2.8, 3.2, and 2.2 above the respective Poisson noise limits. The limiting source of noise in our measurements is not photon statistics, but systematic instrumental and atmospheric effects. Our noise levels are similar to those achieved by other ground-based eclipse observations \citep[e.g.,][also with WIRC]{Zhao2012}. 

{Both $K_S$ band datasets appear to have $>$$3\sigma$ outliers when binned in 12-minute intervals and thus large formal errors. These outliers are not obviously correlated with the seeing conditions, stellar positions, or other easily measured factors. Systematics related to the unstable, highly irregular stellar PSFs, rather than anomalous flux measurements from single images, likely produce these outliers; obtaining a more stable PSF would increase the precision of our eclipse depth measurements.} 

\section{Discussion}
\label{sec:diss}

\subsection{Orbital Eccentricity}

We predicted the timing of the secondary eclipses under the assumption that the planetary orbits are circular, using ephemerides derived from radial velocity and transit observations. The observed time of secondary eclipse may be offset if the brightness distribution on the planet's dayside is non-uniform, as has been observed \citep[e.g.,][]{Knutson2007,Cowan2011}, but the expected time offsets are less than one minute for most hot Jupiters, which is less than the precision of our measurements. A non-zero orbital eccentricity would also alter the timing of the secondary eclipse. From \citet{Charbonneau2005},
\begin{equation}
|e \cos \omega|  \simeq \frac{\pi\Delta t}{2P},
\end{equation}
where $e$ is orbital eccentricity, $\omega$ is the argument of periastron, and $P$ is the orbital period. For HAT-P-23b, $\Delta t$ = 4.0 $\pm$ 2.4~min translates into $|e\cos\omega|$ $\simeq$ 0.0036, constrained to 0.0080 within 2$\sigma$, which is consistent with the estimate of $e\cos\omega \simeq -0.048 \pm 0.023$ from \citet{Bakos2011}. Likewise, for WASP-48b, $\Delta t$ = 2.6 $\pm$ 3.9~min means $|e\cos\omega|$ $\simeq$ 0.0013, constrained to 0.0053 within 2$\sigma$, consistent with the circular orbit assumed in \citet{Enoch2011}. We therefore conclude that, unless the semi-major axes of these orbits are coincidentally aligned with our line of sight, these planets must have effectively circular orbits.

\subsection{Atmospheric Models}

In this section, we combine our secondary eclipse measurements to provide constraints on the shape of the dayside emission spectrum for each planet. For very high signal-to-noise data with broad wavelength coverage, we could, in theory, retrieve the temperature profile for the planetary atmosphere, along with the abundances of chemical species such as CO, CO$_2$, H$_2$O, and CH$_4$ \citep[e.g.,][]{Madhusudhan2009,Line2012,Benneke2012,Lee2012}. Unfortunately, the information content of our secondary eclipse measurements is not high enough to allow for a full retrieval approach \citep{Line2013}.
 
We instead consider three types of atmospheric models. First, we model the planetary emission spectra as blackbodies. In Table~\ref{table}, we report the brightness temperatures, $T_b$, at which the observed eclipse depths equal the planet/star flux ratios averaged over each waveband, where we used synthetic spectra from the new PHOENIX library \citep{Husser2013} for the stars. All of the eclipse depths for WASP-48b and HAT-P-23b can be described by isothermal blackbody emission spectra. {Next, we consider two sets of more complicated atmospheric models. We do not attempt to obtain quantitative best fits using either of the following models, because neither was intended for use in a fitting algorithm and because we cannot tightly constrain the various free parameters \citep[c.f.,][]{Todorov2013}. Instead, we consider a range of models for each planet and qualitatively discuss those that best match our data.}

We compare our data to spectra generated using the methods of \citet{Fortney2008}, which assume a one-dimensional, plane-parallel atmosphere, local thermodynamic equilibrium, solar composition, and equilibrium abundances of chemical species. The absorber TiO is added in equilibrium abundance to the upper atmosphere in some models and excluded from others in order to account for its potential loss due to settling or cold traps. The magnitude of stellar flux incident at the top of the planetary atmosphere is multiplied by a scaling factor to account for the redistribution of energy to the planet's nightside. Specifically, $f=0.5$ signifies even distribution of incident energy over the dayside, but no circulation to the nightside, whereas $f=0.25$ corresponds to total redistribution of energy across the entire planet. 

Finally, we consider spectra generated using the methods of \citet{Burrows2008}, which similarly assume local thermodynamic equilibrium, solar composition, and a plane-parallel atmosphere. These models incorporate a generalized gray absorber in the stratosphere, which is parametrized with an absorption coefficient, $\kappa_e$ (units of cm$^2$~g$^{-1}$), along with a heat sink at a fixed pressure level between 0.01-0.1 bars. The presence of the absorber raises the local atmospheric temperature, producing a high-altitude temperature inversion. A dimensionless parameter $P_n$ represents the efficiency of energy redistribution, with $P_n$ = 0.0 signifying redistribution on the dayside only and $P_n$ = 0.5 representing complete redistribution.

\subsubsection{WASP-48b}

\begin{figure}
\noindent\includegraphics[width=20pc]{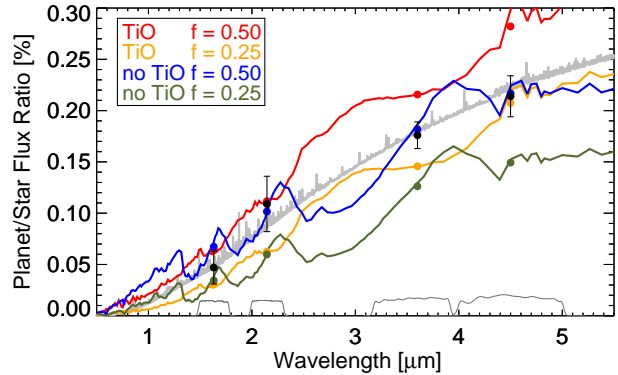}
\caption{Planet/star flux ratio vs. wavelength for four model atmospheres \citep{Fortney2008} for the dayside of WASP-48b. The band-averaged flux ratios are plotted as colored circles. Our secondary eclipse depths are plotted as black circles with 1$\sigma$ error bars. The red and gold models feature TiO in the upper atmosphere, which produces a temperature inversion. The red and blue models were constructed assuming that the planet only radiates from its dayside; the gold and green models represent even radiation from the entire planet. The gray model is the best-fit blackbody emission spectrum. The gray lines at the bottom represent the photometric bands, in arbitrary units.} \label{fig:48_jf}
\end{figure}

\begin{figure}
\noindent\includegraphics[width=20pc]{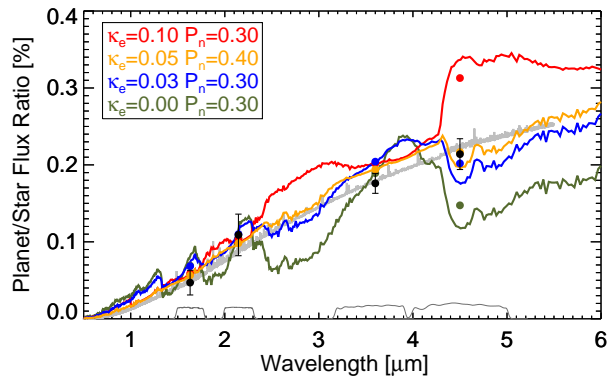}
\caption{Planet/star flux ratio vs. wavelength for four model atmospheres \citep{Burrows2008} for the dayside of WASP-48b. The band-averaged flux ratios are plotted as colored circles. Our secondary eclipse depths are plotted as black circles with 1$\sigma$ error bars. The gray lines at the bottom represent the photometric bands, in arbitrary units. The green, blue, gold, and red models feature a stratospheric absorber with progressively increasing opacity, $\kappa_e$. The gold model has slightly more redistribution of energy from the dayside to the nightside, while the other three models only allow moderate redistribution. The gray model is the best-fit blackbody emission spectrum.} \label{fig:48_ab}
\end{figure}

The predicted equilibrium temperature for WASP-48b is 2400~K if incident energy is re-radiated from the dayside alone and 2000~K if the entire planet radiates evenly, assuming zero albedo. To calculate the brightness temperatures corresponding to our secondary eclipse depths, we used a PHOENIX spectrum for WASP-48 with $T_{eff}$ = 6000~K, $\log g$ = 4.0, and [Fe/H] = 0.0. We find that the best-fit blackbody emission spectrum has $T_{eff}$ = 2158 $\pm$~100~K, with $\chi^2$ = 0.88 {(all our fits for effective temperature have one degree of freedom)}, as shown in Figures~\ref{fig:48_jf} and~\ref{fig:48_ab}.

Figure~\ref{fig:48_jf} compares our secondary eclipse depths to four models of the dayside emission spectrum of WASP-48b following \citet{Fortney2008}. All four eclipse depths are consistent to within $\sim$1$\sigma$ with the model without TiO in the atmosphere and with $f$ = 0.50, indicating little redistribution of energy from the dayside to the nightside. The \textit{Spitzer} measurements provide the strongest constraints because these atmospheric models are most divergent at \textit{Spitzer} wavelengths. The ground-based data are also well-matched by the nominal model preferred by the \textit{Spitzer} data. The model with TiO and $f$ = 0.25 matches the 4.5~$\mu$m and $H$ band measurements and might be consistent with the other two eclipse depths if the amount of recirculation were reduced.

Figure~\ref{fig:48_ab} compares four models generated using the methods of \citet{Burrows2008} to our data. Unlike the models of \citet{Fortney2008}, the band-averaged flux ratios are nearly identical among the different models in the $H$, $K_S$, and 3.6~$\mu$m bands. The eclipse depth at 4.5~$\mu$m, however, is only consistent with two models. In the first model, WASP-48b has a non-zero, but low, abundance of a gray absorber and little redistribution of energy. In the second model, WASP-48b has an intermediate amount of recirculation to the night side and a weak temperature inversion.

Both sets of models indicate that WASP-48b likely has a weak or absent temperature inversion, but the inferred magnitude of energy recirculation depends on the assumed model. Like the simple blackbody fit, the models of \citet{Burrows2008} suggest an intermediate amount of circulation. The best-fit model following \citet{Fortney2008} implies very little recirculation to the planet's nightside, but the gold model with TiO and more efficient day-night circulation also provides a reasonably close match to the data.

\subsubsection{HAT-P-23b}

\begin{figure}
\noindent\includegraphics[width=20pc]{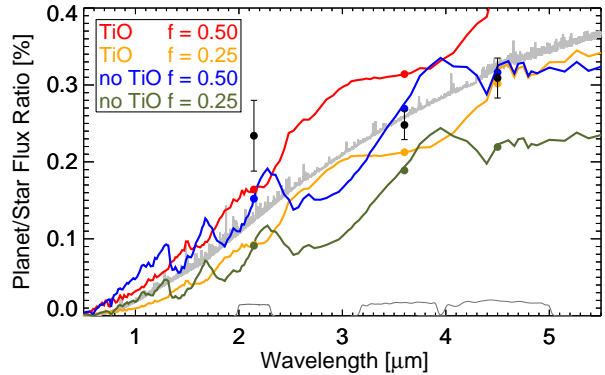}
\caption{Planet/star flux ratio vs. wavelength for four model atmospheres \citep{Fortney2008} for the dayside of HAT-P-23b. The band-averaged flux ratios are plotted as colored circles. Our secondary eclipse depths are plotted as black circles with 1$\sigma$ error bars. The red and gold models feature TiO in the upper atmosphere, which produces a temperature inversion. The red and blue models were constructed assuming that the planet only radiates from its dayside; the gold and green models represent even radiation from the entire planet. The gray model is the best-fit blackbody emission spectrum. The gray lines at the bottom represent the photometric bands, in arbitrary units.} \label{fig:23_jf}
\end{figure}

\begin{figure}
\noindent\includegraphics[width=20pc]{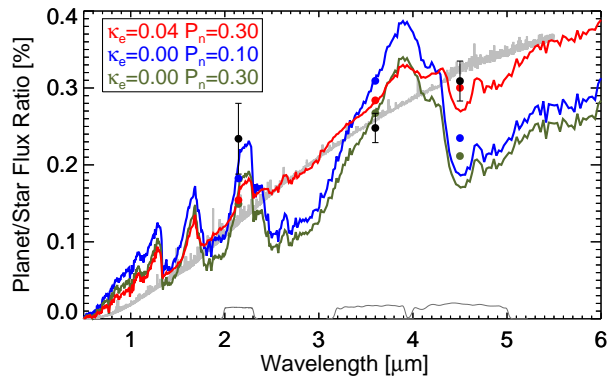}
\caption{Planet/star flux ratio vs. wavelength for three model atmospheres \citep{Burrows2008} for the dayside of HAT-P-23b. The band-averaged flux ratios are plotted as colored circles. Our secondary eclipse depths are plotted as black circles with 1$\sigma$ error bars. The gray lines at the bottom represent the photometric bands, in arbitrary units. The red model features an opaque absorber in the upper atmosphere. The red and green models have moderate redistribution of energy from the dayside to the nightside; the blue model allows very little redistribution of energy. The gray model is the best-fit blackbody emission spectrum.} \label{fig:23_ab}
\end{figure}

The predicted equilibrium temperature for HAT-P-23b is 2450~K if incident energy is re-radiated from the dayside alone and 2050~K if the entire planet radiates evenly, assuming zero albedo. To calculate the brightness temperatures corresponding to our secondary eclipse depths, we interpolated spectra in the PHOENIX library to produce one with $T_{eff}$ = 5900~K, $\log g$ = 4.3, and [Fe/H] = 0.0 for HAT-P-23. The best-fit blackbody emission spectra has $T_{eff}$ = 2154 $\pm$~90~K ($\chi^2$ = 6.11), as shown in Figures~\ref{fig:23_jf} and~\ref{fig:23_ab}. Excluding the secondary eclipse in the $K_S$ band dramatically improves the fit ($\chi^2$ = 0.01), but only lowers $T_{eff}$ to 2123 $\pm$~81~K.

Figure~\ref{fig:23_jf} compares our secondary eclipse depths to four models of the dayside emission spectrum of HAT-P-23b following \citet{Fortney2008}. Again, the \textit{Spitzer} eclipse depths are consistent to within $\sim$1$\sigma$ with the model with $f$ = 0.50 and without TiO in the atmosphere. The $K_S$ band eclipse depth is higher than all of the models, but this band suffers from a higher level of time-correlated noise and the errors are correspondingly large.

Figure~\ref{fig:23_ab} shows three models following \citet{Burrows2008} compared to our secondary eclipse depths. Within $\sim$2$\sigma$, the $K_S$ band eclipse depth is consistent with all three models. The \textit{Spitzer} eclipse depths, particularly the 4.5~$\mu$m band, favor the model with a weak temperature inversion and moderate redistribution of energy, comparable to the preferred model for WASP-48b.

\subsubsection{Stellar Activity and Temperature Inversions}

\newcommand{\caii}{\ion{Ca}{2} H \& K}
\newcommand{\sval}{\ensuremath{S_{\mbox{\scriptsize HK}}}}
\newcommand{\logr}{log(\ensuremath{R'_{\mbox{\scriptsize HK}}})}
\newcommand{\feh}{\ensuremath{[\mbox{Fe}/\mbox{H}]}}

We obtained estimates of the activity levels of both stars using observations of the \caii~emission line strengths from archival Keck HIRES data, following the method described in \citet{Isaacson2010}. For HAT-P-23 we find a median \sval~of 0.344 and a corresponding \logr~value of -4.402, assuming a $B-V$ color of 0.60 calculated using the star's effective temperature and stellar atmosphere models. For WASP-48, we obtained a single measurement with a \sval~of 0.139 and \logr~equal to -5.135, assuming a $B-V$ color of 0.57 calculated from stellar atmosphere models. We find that HAT-P-23's elevated activity level is consistent with our hypothesis that increasing the UV flux suppresses the formation of temperature inversions by destroying the absorber responsible for the inversions \citep{Knutson2010}. WASP-48b would appear to be an exception to this rule, as the planet appears to have a weak or absent inversion very similar to that of HAT-P-23b but is found orbiting a quiet star. In this case, it is more likely that the planet never contained the right elemental abundances to form the absorbing molecule in the first place, as proposed by \citet{Madhusudhan2012} for the case of TiO and planets with high C/O ratios.  It is worth noting that WASP-48b orbits what may be a modestly metal-poor star (\feh~of $-0.12\pm0.12$, \citealp{Enoch2011}), while HAT-P-23b orbits a metal-rich star (\feh~of $+0.15\pm0.04$, \citealp{Bakos2011}). If WASP-48b's atmosphere has a lower metallicity than that of HAT-P-23b, it could explain the absence of a strong temperature inversion.  

\section{Conclusions}

We observed secondary eclipses of the extrasolar hot Jupiters WASP-48b and HAT-P-23b. In the $H$, $K_S$, 3.6~$\mu$m, and 4.5~$\mu$m bands, respectively, we measured secondary eclipse depths of 0.047 $\pm$ 0.016\%, 0.109 $\pm$ 0.027\%, 0.176 $\pm$ 0.013\%, and 0.214 $\pm$ 0.020\% for WASP-48b. In the $K_S$, 3.6~$\mu$m, and 4.5~$\mu$m bands, respectively, we measured secondary eclipse depths of 0.234 $\pm$ 0.046\%, 0.248 $\pm$ 0.019\%, and 0.309 $\pm$ 0.026\% for HAT-P-23b. The dayside emission spectra of both planets are well-described by simple blackbody models with effective temperatures of 2158 $\pm$~100~K for WASP-48b and 2154 $\pm$~90~K for HAT-P-23b. They also match models generated using the methods of \citet{Fortney2008} in which TiO is absent from their atmospheres and the efficiency of energy transport from their daysides to nightsides is low. Models generated using the methods of \citet{Burrows2008} tend to prefer weak temperature inversions and moderate day/night recirculation. The low measured timing offsets from the predicted centers of secondary eclipse for WASP-48b and HAT-P-23b constrain $|e\cos\omega|$ to 0.0053 and 0.0080 within 2$\sigma$, respectively, suggesting that these planets likely have circular orbits.

Our results demonstrate the continued ability of warm \textit{Spitzer} to characterize the atmospheres of hot Jupiters, as well as the utility of expanding this wavelength coverage with complementary ground-based measurements. Although our near-IR $H$ and $K_S$ band measurements were consistent with the majority of the models considered here, they serve as an additional confirmation that the assumptions used in these models, including solar composition atmospheres, provide a reasonably accurate description of these planetary atmospheres. The installation of a new diffusing filter in the WIRC filter wheel in fall 2013 should allow us to achieve a much more stable and evenly distributed PSF, mitigating many of the issues encountered with the data presented here. If future observations are able to lower the systematic noise floor to a level approaching the photon noise limit instead of the factor of $\sim$2-3 achieved here, these observations will play a crucial role in extending and refining the knowledge gained from \textit{Spitzer} secondary eclipse observations in the 3.6 and 4.5~$\mu$m bands. This will provide invaluable constraints for more in-depth modeling, including scenarios with non-equilibrium chemistries, higher metallicity atmospheres, and super- or sub-solar C/O ratios.

\acknowledgments

J.G.O. receives support from the National Science Foundation's Graduate Research Fellowship Program and thanks Michael Line for many helpful discussions. J.-M.D. acknowledges funding from NASA through the Sagan Exoplanet Fellowship program administered by the NASA Exoplanet Science Institute (NExScI). M.Z. is supported by the Center for Exoplanets and Habitable Worlds at Penn State University, and the AAS Small Research Grant. This work is based on observations made with the \textit{Spitzer Space Telescope}, which is operated by the Jet Propulsion Laboratory, California Institute of Technology, under contract with NASA. The Palomar Hale Telescope is operated by Caltech, JPL, and Cornell University. We thank the anonymous referee for many helpful comments.


\end{document}